\documentclass[pdflatex,sn-aps]{sn-jnl}

\usepackage{graphicx}%
\usepackage{multirow}%
\usepackage{amsmath,amssymb,amsfonts}%
\usepackage{amsthm}%
\usepackage{mathrsfs}%
\usepackage[title]{appendix}%
\usepackage{xcolor}%
\usepackage{textcomp}%
\usepackage{manyfoot}%
\usepackage{booktabs}%
\usepackage{algorithm}%
\usepackage{algorithmicx}%
\usepackage{algpseudocode}%
\usepackage{listings}%

\theoremstyle{thmstyleone}%
\theoremstyle{thmstyletwo}%

\theoremstyle{thmstylethree}%

\begin{document}

\title[The thermodynamics of liquid-vapor coexistence for a van der Waals fluid. Analytical solution of the Clausius-Clapeyron equation]{The thermodynamics of liquid-vapor coexistence for a van der Waals fluid. Analytical solution of the Clausius-Clapeyron equation}


\author*[1]{\fnm{J. L.} \sur{Cardoso}}\email{jlcc@azc.uam.mx}

\author[1]{\fnm{V. G.} \sur{Ibarra-Sierra}}\email{vgis@azc.uam.mx}
\equalcont{These authors contributed equally to this work.}

\author[1]{\fnm{J. C.} \sur{Sandoval-Santana}}\email{jcss@azc.uam.mx}
\equalcont{These authors contributed equally to this work.}

\author[1]{\fnm{A.} \sur{Kunold}}\email{akb@azc.uam.mx}
\equalcont{These authors contributed equally to this work.}

\affil*[1]{\orgdiv{\'Area de F\'isica Te\'orica y Materia Condensada}, \orgname{Universidad Aut\'onoma Metropolitana, Azcapotzalco}, \orgaddress{\street{Av. San Pablo Xalpa 180}, \city{Ciudad de M\'exico}, \postcode{02200}, \state{Ciudad de M\'exico}, \country{M\'exico}}}


\abstract{This work presents a pedagogical derivation of
the thermodynamics of a van der Waals fluid by explicitly incorporating
pairwise molecular interactions and the finite size of particles into
the statistical-mechanical description. Starting from the Lennard–Jones
potential, we evaluate the second virial coefficient to infer the 
virial expansion of the equation of state and recover the van der Waals 
equation using only its leading correction. The corresponding partition
function allows us to obtain all thermodynamic potentials for both 
monoatomic and diatomic fluids in a transparent and instructive manner.

Building on this framework, we formulate and solve analytically the 
Clausius–Clapeyron equation in the vicinity of the critical point,
obtaining the liquid-vapor coexistence curve in closed form. This 
approach not only clarifies the microscopic origin of van der Waals 
thermodynamics but also complements-and in several aspects
improves upon-traditional treatments that rely heavily on numerical methods or 
heuristic arguments.

In addition, because the van der Waals equation naturally predicts the 
liquid–vapor equilibrium, the existence of critical points, and the 
functional form of the saturation curve
of the pressure as a function of temperature,
it provides an 
analytically tractable framework for studying a 150-year-old problem 
that has historically been addressed using graphical constructions or
numerical solutions. As such, the formulation developed here offers a 
coherent, accessible, and conceptually unified route for students and 
instructors to understand phase coexistence in simple fluids from first 
principles.
}




\maketitle

\section{Introduction}\label{sec1}

The van der Waals equation of state (VDWES),
originally proposed in 1873 \cite{vanderwaals}
and later examined in depth by Barker \cite{JABarker}, remains a
paradigmatic model for understanding non-ideal fluids and their phase
transitions. Classic treatments \cite{callen,huang,atkins}
emphasize that, despite its apparent simplicity, it successfully captures
key thermodynamic features such as the liquid-vapor transition and the
existence of a critical point. Yet, from a modern perspective,
its traditional derivation is not rigorous: it relies on a mean-field
picture in which monoatomic molecules behave as hard spheres that reduce
the available volume per particle, while an average attractive interaction
produces a cohesive pressure proportional to $v^{-2}$.

Although widely presented in textbooks, the derivation of the VDWES
is often heuristic or purely phenomenological. This motivates a more
transparent and fully microscopic development. One of the aims of this
work is therefore to provide a pedagogical derivation based on the virial
expansion and the Lennard-Jones intermolecular potential, following the
foundational statistical-mechanical formalisms of Mayer \cite{mayer},
Hirschfelder et al. \cite{hirschfelder}, Pathria and Beale \cite{pathria},
and Huang \cite{huang}. This approach makes explicit the physical origin
of the corrective terms that distinguish the VDWES from the ideal gas
equation of state (IGES): one term accounting for the finite size of
molecules, which effectively reduces the accessible volume,
and another describing attractive forces, which contribute
an additional pressure.

Just as the thermodynamic properties of an ideal gas follow directly
from its partition function, the corresponding potentials of a van der
Waals fluid (VDWF) can also be obtained once the microscopic approximation
for the partition function is specified. However, standard treatments rarely
explore this route in detail, and as a consequence they often fail to
connect the resulting thermodynamic potentials with the phenomenology
of phase transitions. In this work we revisit this connection and
show how the approximations leading to the VDWES naturally yield the
full set of thermodynamic potentials of the VDWF.

A second objective of this study concerns liquid–vapor coexistence.
Analytical solutions of the Clausius-Clapeyron equation for the
van der Waals coexistence curve are uncommon.
Existing results provide valuable approximations,
while classical discussions \cite{JLekner} rely heavily
on graphical or numerical approaches.
By leveraging the explicit form of the thermodynamic potentials
derived here, we obtain an analytical solution of the Clausius-Clapeyron
equation and express the coexistence curve in closed form.
This stands in contrast to earlier studies, which typically resorted to
numerical methods
\cite{JABarker,WGChapman,JGPowles,JLekner,KJRunge,LSIgalotti}
due to the complex coupling of variables in the VDWES.
The resulting framework thus offers both pedagogical clarity
and an analytical treatment of phase coexistence within the van der Waals model.

\section{The van ver Waals equation of state and its thermodynamic implications}

The equation of state proposed by J. D. van der Waals in 1873
\cite{vanderwaals, atkins, callen}
as part of his doctoral thesis can be written as
\begin{equation}
P = \frac{nRT}{V - nb} - a\left(\frac{n}{V}\right)^2 = \frac{RT}{v - b} - \frac{a}{v^2},
\label{EcEdo:VanDerWaals}
\end{equation}
where $P$ is the pressure, $T$ the temperature, $V$ the volume, $n$ the number
of moles ($n = N/N_A$, with $N$ the number of particles and $N_A$ Avogadro’s number),
and $R$ the universal gas constant. The parameter $b$ represents the effective
volume excluded per mole of particles, while $a$ quantifies the strength of the
intermolecular attractive forces.
Compared with the ideal gas equation of state, the van der Waals equation
introduces two corrective terms: a volume correction $(V - nb)$ accounting
for the finite size of molecules, and a pressure correction $-a(n/V)^2$
that incorporates the effect of intermolecular attractions.

\subsection{Critical point and the law of corresponding states}

To determine the critical point, the equation of state (\ref{EcEdo:VanDerWaals}) is rewritten as a cubic polynomial in 
terms of $v$
\begin{equation}
v^3 -\left(\frac{RT}{P}+b\right)v^2 +\frac{a}{P} v -\frac{ab}{P} =0. 
\label{Polinomio3}
\end{equation}
The resulting cubic equation for the molar volume can have either one or three real
roots, depending on the values of temperature and pressure.
Since each real root corresponds to a distinct thermodynamic state, we are only interested
in these physical (real) solutions. At very high temperatures, only one real root exists,
corresponding to the gaseous phase. As the temperature decreases, within a certain range
of pressures, three real roots appear; between the smallest and largest of these,
the system exhibits liquid–vapor coexistence. There is a particular state at which
the transition from one to three real roots occurs—this is the critical point,
characterized by the critical values 
$\left(P_c, T_c, v_c\right)$
\cite{guggenheim},
where the van der Waals equation takes the form
\begin{eqnarray}
\left(v-v_c\right)^3 =  v^3 -3v_c v^2 +3v_c^2 v -v_c^3 =0 .
\label{Polinomio3:crit}
\end{eqnarray}
Table~\ref{tab:fluidsdata} presents the critical values for
several representative molecules.
From the above relations, the critical point can be determined,
and consequently, the parameters $a$ and $b$
can be calculated as
\begin{eqnarray}
a &=& 3 P_c v_c^2, \\
b &=& \frac{v_c}{3}, \\
T_c &=& \frac{8 P_c v_c}{3 R}.
\end{eqnarray}
By substituting these parameters into the van der Waals equation
of state (VDWES) and expressing it in terms of the critical variables, we obtain
\begin{equation}
p_r = \frac{8 t_r}{3 v_r - 1} - \frac{3}{v_r^2}.
\label{EcEdo:reducida}
\end{equation}
In this equation, the reduced variables
$p_r = P/P_c$, $t_r = T/T_c$, and $v_r = v/v_c$
have been introduced to simplify the van der Waals equation.
This formulation allows different substances to be characterized by
their corresponding critical points through a single, universal
equation of state-a principle known as the law of corresponding states.
The reduced VDWES [Eq.~\eqref{EcEdo:reducida}] implies that, regardless
of the substance, each isotherm plotted as $p_r$ versus $v_r$
represents the behavior of any fluid.
Even though in absolute values their individual critical points differ,
in all cases, the reduced critical point is located at 
$\left(p_r,v_r,t_r\right) = \left(1,1,1\right)$.

\begin{figure}[h]
\centering
\includegraphics[width=0.9\textwidth]{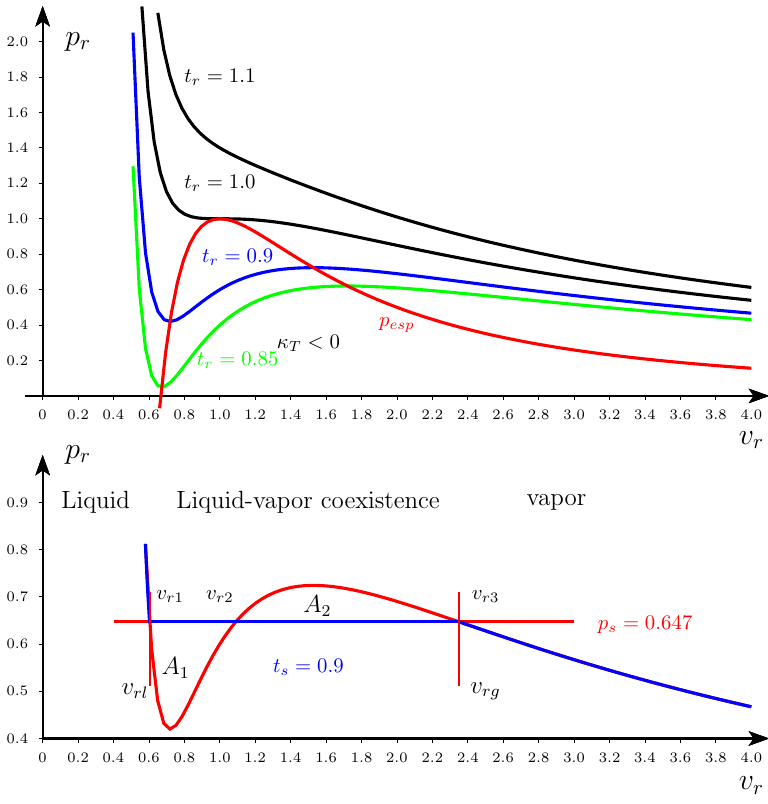}
\caption{Isotherms according to the van der Waals equation of state.
In the upper graphs, one isotherm corresponds to $t_r > 1$,
another to $t_r = 1$, and two others to $t_r < 1$.
In addition, the spinodal pressure is plotted.
Note that $p_{\mathrm{esp}}$ passes through the minima and maxima of
the isotherms for $t_r \leq 1$. Plotting the spinodal pressure is useful
because it delineates the unstable thermodynamic states,
since below this curve $\kappa_T$ becomes negative.
The procedure to determine the vapor pressure, according to Maxwell’s rule,
is shown in the lower panel.
It consists of replacing the section of the isotherm between $v_{r1}$
and $v_{r3}$ with a horizontal line such that the area between this
line and the curve from $v_{r1}$ to $v_{r2}$, $A_1$, is equal to the area
between the curve from $v_{r2}$ to $v_{r3}$ and the same line, $A_2$.}
\label{IsoVanDerWaals}
\end{figure}

\begin{table}
\caption{
Critical values for
various monatomic and diatomic fluids \cite{callen}.}
\label{tab:fluidsdata}
\begin{tabular}{|l|r|r|r|c|c|}
\hline
Molecule & Mass  & $T_c (K)$ & $P_c (atm)$ & $a$ &
$b$ \\
 & (g/mol) & & & $(10^6 atm cm^6/mol^2)$ & $(cm^3/mol)$ \\ \hline
He$^3$ & 3.00 & 3.309 & 1.124 & 0.027674 & 30.19 \\ 
He$^4$ & 4.00 & 5.189 & 2.260 & 0.033844 & 23.55 \\
Ne & 20.17 & 44.500 & 26.000 & 0.220142 & 17.74 \\
Ar & 39.94 & 150.720 & 47.990 & 1.344740 & 32.21 \\ 
Kr & 83.80 & 290.400 & 54.230 & 2.297000 & 39.60 \\
Xe & 131.20 & 289.730 & 57.580 & 4.141540 & 51.61 \\
\hline
F$_2$ & 37.99 & 144.3 & 51.42 & 1.150400 & 28.78 \\
Cl$_2$ & 70.00 & 416.9 & 78.65 & 6.277860 & 54.37 \\
Br$_2$ & 79.90 & 588.0 & 101.5 & 9.676880 & 54.42 \\
I$_2$ & 126.9 & 819.0 & 116.0 & 16.42690 & 72.42 \\
\hline
\end{tabular}
\end{table}
The interpretation of an isotherm in a $p_r-v_r$
diagram requires the expression for the isothermal compressibility
coefficient of a van der Waals fluid (VDWF), given by
\begin{equation}
\kappa_T = -\frac{1}{v}\left(\frac{\partial v}{\partial P}\right)_T
= \frac{1}{P_c} \frac{v_r -1/3}{v_r} \frac{1}{p_r -p_{esp}},
\end{equation}
where
\begin{equation}
p_{esp} = \frac{3v_r-2}{v_r^3} 
\end{equation}
is the spinodal pressure.
Figure \ref{IsoVanDerWaals} shows several isotherms in a
$p_r-v_r$ diagram. One isotherm corresponds to $t_r>1$, another to $t_r=1$,
and the remaining two to $t_r<1$.
The spinodal pressure is also plotted. Note that $p_{esp}$ passes through the 
minima and maxima of the isotherms for $t_r\leq1$. Plotting the spinodal 
pressure is useful because it delineates unstable thermodynamic states,
where the isothermal compressibility 
$\kappa_T$ becomes negative below this curve.

Thomas Andrews, years before van der Waals, conducted a series
of experiments to study the liquid–vapor transition of $\text{CO}_2$ \cite{andrews}.
His experimentally measured isotherms showed that for $t_r < 1$,
the liquid–vapor phase change occurs at constant temperature and pressure;
that is, the isotherms exhibit a plateau at a specific pressure value,
known as the saturation pressure or vapor pressure.
This horizontal line starts at a value $v_{rl}$, called the liquid volume,
and ends at the gas (or vapor) volume $v_{rg}$.
Both phases coexist in the region between $v_{rl}$ and $v_{rg}$.
Andrews determined that the saturation pressure increases with
temperature, while the volume difference, $\Delta v = v_{rg} - v_{rl}$, decreases.
The volume difference vanishes at the critical temperature.

The VDWES has been used to explain the behavior of isotherms for
different substances. If a pressure value is chosen between the
maximum and minimum of the isotherm ($t_r < 1$), three roots,
$v_{r1}$, $v_{r2}$, and $v_{r3}$,
are obtained, where $v_{r1} < v_{r2} < v_{r3}$.
These correspond to the positions of the different phases in the
$p_r$–$v_r$ diagram.
The liquid and vapor phases are located at $v_l = v_{r1}$
and $v_g = v_{r3}$, respectively.
To select the appropriate pressure, Maxwell’s rule is applied.
This rule consists of replacing the section of the isotherm between
$v_{r1}$ and $v_{r3}$ with a horizontal line such that the area between
this line and the curve from $v_{r1}$ to $v_{r2}$
is equal to the area between the curve from $v_{r2}$ to $v_{r3}$
and the same line.
The above is justified because, in a reversible process,
a substance must absorb or release a certain amount of heat
to undergo a phase change, regardless of the path by which
the process is carried out (see Fig.~\ref{IsoVanDerWaals}).

\subsection{Other thermodynamic variables}

Using the first law of thermodynamics together with
the Maxwell relations \cite{callen, fermi}, one obtains the following
relation between the molar internal energy $u=U/n$,
and the VDW equation of state:
\begin{equation}
\left(\frac{\partial u}{\partial v}\right)_T
= T\left( \frac{\partial P}{\partial T} \right)_v - P.
\label{Rel:EcEdos}
\end{equation}
This identity implies that the internal energy \cite{GreinerThermoStatMech} has the form
\begin{equation}
u = u_0 + c_v T - \frac{a}{v},
\end{equation}
where $c_v$ is the molar heat capacity at constant
volume and $u_0$ is an integration constant.
The corresponding molar entropy is
\begin{equation}
s = s_0 + R \ln\left[ T^{c_v/R}\left(v - b\right) \right].
\end{equation}
Here we assume that the heat capacity is constant,
which is valid at sufficiently high temperatures.

By applying the corresponding Legendre transformations \cite{GreinerThermoStatMech},
the other molar thermodynamic potentials \cite{fermi} are obtained as
\begin{eqnarray}
h &=& u_0 +RT\left[\frac{c_v}{R} +\frac{v}{v-b}\right]-2\frac{a}{v}, \\
f &=& u_0  -RT\left\{\frac{s_0-c_v}{R} 
+\ln\left[T^{c_v/R}\left(v-b\right)\right]\right\}-\frac{a}{v}, \\
g &=& u_0 
-RT\left\{\frac{s_0-c_v}{R} -\frac{v}{v-b}
+\ln\left[T^{c_v/R}\left(v-b\right)\right]\right\} 
-2\frac{a}{v}.
\end{eqnarray}
Here, $h$ denotes the molar enthalpy, 
$f$ the molar Helmholtz free energy, and 
$g$ the molar Gibbs free energy.
Although these potentials can be obtained straightforwardly,
they are not particularly useful for describing a VDW fluid.
In particular, the molar Gibbs free energy depends on both volume 
$v$ and temperature $T$,
making it difficult to express solely in terms of 
$P$ and $T$, which are its natural variables.

\section{Statistics of a van der Waals fluid}

In this section, we will derive the partition function
of a van der Waals fluid using the second virial coefficient,
which can be calculated from the Lennard–Jones
intermolecular potential \cite{LennardJonesPotential} (or a similar potential)
for monatomic and diatomic particles.

\subsection{Monoatomic fluid}

The fluid we are going to analyze is no longer diluted,
so the interaction of its particles will 
occur in a binary form, given the possible proximity between
monatomic molecules. The 
Hamiltonian is written as \cite{pathria, GreinerThermoStatMech}
\begin{equation}
H\left(\vec q, \,\vec p\right) =\sum_{i=1}^{3N} \frac{p_i^2}{2m} +\sum_{i<j} 
\phi_{ij}\left(\vec q_{ij}\right), 
\end{equation}
where $\phi_{ij}\left(\vec q_{ij}\right)$ is the intermolecular
potential and $\vec q_{ij}=\vec q_j -\vec q_i$
is the relative coordinate of particles $i$ and $j$, with 
$i,j = 1,2,\dots N$.
Since we are analyzing a fluid near the liquid–vapor transition,
the temperature is sufficiently high for the classical approximation
to hold; therefore, the problem will not be treated using quantum mechanics.

\begin{figure}[h]
\centering
\includegraphics[width=0.7\textwidth]{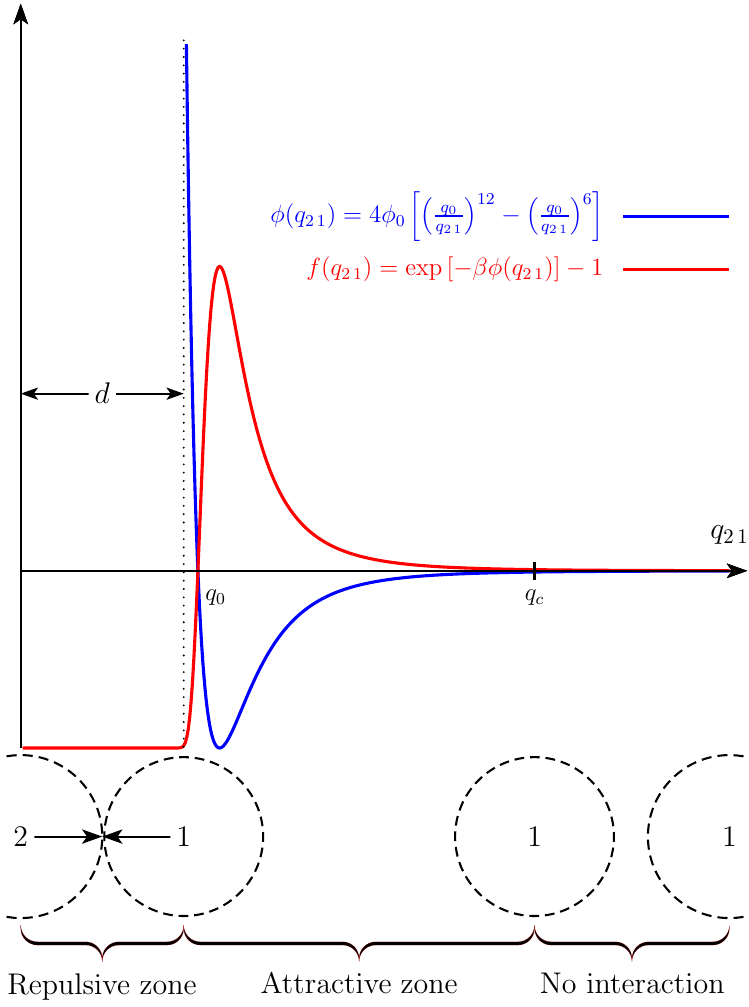}
\caption{The Lennard–Jones intermolecular potential
and the corresponding Mayer function are shown.
The blue curve illustrates the Lennard–Jones potential,
while the red curve represents the Mayer function evaluated
with $\phi\left(q_{2\,1}\right)$.
Spheres representing monatomic molecules are depicted
below; if these spheres are taken to be hard,
two regions appear: a repulsive region for
$q\leq q_0$ and a short-range attractive region
beyond this point. Within the repulsive region, the Mayer
function exhibits an inflection point at 
$d$, which is interpreted as the diameter of the hard spheres.}
\label{Lennard-Jones}
\end{figure}

The most commonly used intermolecular potential
describing the interaction between particles
is the Lennard–Jones potential \cite{LennardJonesPotential}
\begin{equation}
\phi_{ij}\left(\vec q_{ij}\right) = \phi_{ij}\left(q_{ij}\right) =
4\phi_0\left[\left(\frac{q_0}{q_{ij}}\right)^{12} -\left(\frac{q_0}{q_{ij}}\right)^6\right].
\end{equation}
Here $q_0$ is the coordinate where the potential vanishes,
$\phi\left(q_{ij}=q_0\right)=0$, and $-\phi_0$ is the
depth of the attractive well.
Note that this potential is central, since it 
only depends on the magnitude of the relative coordinate
$q_{ij} = \left|\vec q_{ij}\right|$. 
The first term, $4\phi_0\left(q_0/q_{ij}\right)^{12}$,
represents the repulsive part of the potential, while the 
second term, $-4\phi_0\left(q_0/q_{ij}\right)^6$,
corresponds to the attractive interaction,
which arises when the molecules behave as
electric dipoles at short distances.
This attractive contribution is commonly known as
the van der Waals interaction.
The form of the Lennard-Jones potential is shown in 
Fig. \ref{Lennard-Jones}. The function
$\phi _{ij}\left(q_{ij}\right)$ is repulsive for 
$q_{ij}\leq q_0$ and attractive for $q_{ij} >q_0$.

To determine the thermodynamic properties of the interacting fluid,
we work within the canonical ensemble,
where the system is characterized by a fixed number
of particles $N$, volume $V$, and temperature $T$. 
In this ensemble, all macroscopic quantities are
derived from the canonical partition function 
$Q_N(V,T)$, defined as the integral over the
classical phase space of the Boltzmann factor 
$\exp(-\beta H)$.
Using the Hamiltonian introduced above, the partition function
\cite{GreinerThermoStatMech,pathria} becomes
\begin{equation}
Q_N(V,T) = \frac{1}{N!h^{3N}}
\int \exp\left[-\beta\left(\sum_{i=1}^{3N}\frac{p_i^2}{2m}
+\sum_{i<j}\phi_{ij}(q_{ij})\right)\right] d^{3N}p d^{3N}q.
\end{equation}
In order to analyze diatomic molecules or molecules
with more complex structures, it is necessary to incorporate
the corresponding degrees of freedom into both the kinetic
energy and the intramolecular potential that couples the
constituent atoms.
However, for monatomic molecules, the integration over
the linear momenta can be carried out directly, yielding
\begin{multline}
Q_N\left(V,T\right) = \frac{1}{N!}\left(\frac{2\pi mkT}{h^2}\right)^{3N/2} 
\int \exp\left[-\beta\sum_{i<j}\phi_{ij}\right] d^{3N}q \\
= \frac{1}{N!\lambda^{3N}} Z_N\left(V,T\right),
\label{Ec:PartitionFunctionTwo}
\end{multline}
where $\lambda = h/\sqrt{2\pi mkT}$ is the de Broglie
thermal wavelength and
\begin{equation}
Z_N\left(V,T\right) = \int \exp\left[-\beta\sum_{i<j}\phi_{ij}\right]d^{3N}q
\end{equation}
is the part of the partition function that depends
on the potential energy of interaction between the
particles. Note that if $\phi_{ij}=0$,
then $Z_N\left(V,T\right) = V^{N}$,
and we recover the partition function
of a monatomic ideal gas.
On the other hand, the integrand can be 
written as
\begin{equation}
e^{-\beta\sum_{i<j}\phi_{ij}} = \prod_{i<j}e^{-\beta\phi_{ij}}.
\end{equation}
To integrate it, the 
Mayer function \cite{pathria,GreinerThermoStatMech}
is introduced
\begin{equation}
f_{ij}\left(q_{ij}\right) = e^{-\beta\phi_{ij}}-1.
\end{equation}
Figure \ref{Lennard-Jones} shows the plot of
the Mayer function obtained from the Lennard–Jones
intermolecular potential.
This function exhibits an inflection point $d<q_0$,
within the repulsive region, which allows us to identify an
impenetrable spherical core when
$f_{ij}\left(q_{ij}\right) =-1$
and the attractive region when $f_{ij}\left(q_{ij}\right)>-1$. 
if we return to the hard-sphere model for monatomic molecules,
we may interpret $d$ as the diameter of the spheres.
In the attractive region, the Mayer function reaches
a maximum and then rapidly decays to zero as
$\phi\left(q_{ij}\right)$ tends to zero.
This confirms that the Lennard–Jones interaction
is short-ranged: molecules separated by several multiples
of $d$ no longer interact significantly.
This cutoff distance is denoted by $q_c$
and defines a cutoff radius, since for $q_{2\,1}>q_c$
the intermolecular potential becomes very weak and
$f_{ij}\left(q_{ij}\right)$ approaches zero exponentially.
Using the Mayer function, the integrand can be expanded as
\begin{multline}
\prod_{i<j}e^{-\beta\phi_{ij}} = \prod_{i<j} \left(1+f_{ij}\right) \\
= \left(1+f_{1\,2}\right)\left(1+f_{2\,3}\right)\left(1+f_{3\,4}\right)\dots \\
= 1+\sum_{i<j}f_{ij} 
+ \sum_{i<j} \sum_{i^\prime<j^\prime} f_{ij}f_{i^\prime j^\prime} 
+\dots
\end{multline}
and write $Z_N\left(V,T\right)$ accordingly
\begin{multline}
Z_N\left(V,T\right) 
= \int \left\{1+\sum _{i<j}f_{ij} + \sum _{i<j} \sum _{i^\prime<j^\prime} 
f_{ij}f_{i^\prime j^\prime} +\dots\right\} d^{3N}q \\
= V^N +\sum_{i<j} \int f_{ij} d^{3N}q
+\sum_{i<j} \sum_{i^\prime<j^\prime} \int f_{ij}f_{i^\prime j^\prime} d^{3N}q
+\dots \label{Z_N}
\end{multline}
The first term corresponds to the monatomic ideal gas.
The second term accounts for corrections arising from binary
interactions, meaning that particles interact only
in pairs and, once a pair is formed, no additional molecule
interacts with that same pair.
The third term represents correlations involving two binary
interactions, allowing for the possibility of
three- or four-particle clusters.
Together, these contributions constitute the virial
expansion of the equation of state. 

In this section, we will integrate the first correction
in order to obtain the van der Waals equation of state
and its associated thermodynamics.
The symbol $\sum_{i<j}$ denotes a double
sum over the indices $i$ and $j$ with $i\neq j$,
this restriction ensures that each distinct pair is counted only once.
Thus, the first correction is
\begin{equation}
\sum_{i<j} \int f_{ij} d^{3N} q 
= \frac{1}{2} \underbrace{\sum_{i=1}^N\sum_{j=1}^N}_{i\neq j} 
\int f_{ij} d^3q_1 d^3q_2 \dots d^3q_N.
\end{equation}
Since the particles are indistinguishable,
we may perform the integration over $f_{1\,2}$;
this is equivalent to integrating over
$f_{2\,3}$, $f_{3\,4}$ or any other Mayer function
corresponding to a pair of molecules.
In this sense
\begin{multline}
\sum_{i<j} \int f_{ij} d^{3N} q 
= \frac{1}{2} N\left(N-1\right) \int f_{1\,2} d^3q_1 d^3q_2 
\underbrace{\int d^3q_3 \dots \int d^3q_N}_{=V^{N-2}}\\
= \frac{V^{N-2}}{2} N^2 \int f_{1\,2} d^3q_1 d^3q_2, 
\end{multline}
since the double sum reduces to $N\left(N-1\right)/2$
equivalent terms, which may be approximated as
$N^2$ for $N>>1$. Several of the integrals, such as those over
$d^3q_3$,contribute simply a factor of the system volume.
To integrate $f_{1\,2}$ we rewrite the integral in terms
of the relative coordinate $\vec q_{1\,2} = \vec q_2 -\vec q_1$
and the center-of-mass coordinate
$\vec q_{cm}=\left(\vec q_2 +\vec q_1\right)/2$.
With this, we obtain
\begin{equation}
\sum_{i<j} \int f_{ij} d^{3N} q = \frac{N^2V^{N-2}}{2} \int f_{1\,2} d^3q_{1\,2} 
\underbrace{\int d^3q_{cm}}_{=V} =  \frac{N^2V^{N-1}}{2} \int f_{1\,2} d^3q_{1\,2} .
\end{equation}
The integral over the center-of-mass coordinates
contributes an additional volume $V$ to
the thermodynamic system.
The integral over the relative coordinates can be evaluated
in spherical coordinates, which leads to the definition
of the second virial coefficient,
\begin{equation}
B\left(T\right) = -2\pi\int_0^\infty f\left(q\right)q^2 dq, \label{B(T)}
\end{equation}
and allows us to express the partition function (\ref{Ec:PartitionFunctionTwo}) as
\begin{equation}
Q_N\left(V,T\right) = \frac{V^N}{N! \lambda^{3N}} \left[1 -\frac{N^2}{V}B\left(T\right)\right].
\end{equation}
It is clear that the second virial coefficient
depends on the temperature only through the
Mayer function. The Helmholtz free energy is defined as
\begin{equation}
F\left(T,V,N\right) = - k T \ln Q_N\left(V, T\right).
\end{equation}
This allows us to calculate the thermal equation of state.
Taking into account that 
$N^2B\left(T\right)/V <<1$, we obtain that
\begin{equation}
P=-\left(\frac{\partial F}{\partial V}\right)_{T,N}=kT\frac{N}{V}\left[1+\frac{N}{V}B\left(T\right)\right] \label{PV:B(T)}
\end{equation}
If the additional corrections of (\ref{Z_N}) are considered,
the higher-order virial coefficients must be estimated,
leading to the equation of state
\begin{equation}
P=kT\frac{N}{V}\left[1+\frac{N}{V}B\left(T\right)+\frac{N^2}{V^2}C\left(T\right)
+\frac{N^3}{V^3}D\left(T\right)+\dots\right]
\end{equation}
Defining and evaluating higher-order virial terms, such as
$C\left(T\right)$ and $D\left(T\right)$, is 
beyond the scope of this work. Here, we will only evaluate
$B\left(T\right)$ to derive the 
VDWES.

Based on the shape of the Meyer function shown in Figure \ref{Lennard-Jones}, the second
virial coefficient 
$B\left(T\right)$ can be evaluated by integrating it as follows
\begin{multline}
B\left(T\right) = -2\pi \int_0^\infty f\left(q\right) q^2dq 
= 2\pi \int_0^d q^2dq 
-2\pi\int_d^\infty 
\underbrace{\left[e^{-\beta\phi\left(q\right)}-1\right]}_{=-\beta\phi\left(q\right)}q^2dq \\
= \frac{2}{3}\pi d^3 +\frac{2\pi}{kT}\int_d^\infty \phi\left(q\right) q^2dq .
\end{multline}
The definition of $B\left(T\right)$ is used in the first step
of the integral.
Then, the repulsive region between the
molecules is separated from the attractive region,
and the corresponding integrals are evaluated.
Finally, since high temperatures are considered, or the intermolecular potential is weak,
we have $\phi(q)<<kT$, and therefore the first-order approximation
of the exponential function can be applied.
With these considerations, the second virial coefficient becomes
\begin{equation}
B\left(T\right) = b^\prime -\frac{a^\prime}{kT}, \label{B(T)-ab}
\end{equation}
where $b^\prime = 2\pi d^3/3$ and
\begin{equation}
a^\prime = -2\pi \int_d^\infty \phi\left(q\right) q^2dq.
\end{equation}
when it is modeled as a rigid sphere.
The form of the second virial coefficient also shows that,
at very high temperatures, $B\left(T\right)$ approaches a constant.
The equation of state then becomes
\begin{multline}
P = kT\frac{N}{V} +\frac{N^2}{V^2} \left[b^\prime -\frac{a^\prime}{kT}\right] kT 
= kT\frac{N}{V} \left[1+\frac{N}{V}b^\prime\right] -a^\prime \frac{N^2}{V^2}  \\
= kT\frac{N}{V} \left[1-\frac{N}{V}b^\prime\right]^{-1} -a^\prime \frac{N^2}{V^2} .
\end{multline}
In the first step, the expression for $B\left(T\right)$
is substituted into (\ref{PV:B(T)}).
In the second step, the term $NkT/V$ is factored out.
Finally, assuming that $Nb^\prime/V <1$, the
geometric series expansion is used to rewrite
the equation of state.
This last equality contains the factor $N/V$,
which can be rewritten using the microscopic
definition of the number of moles
\begin{equation}
\frac{N}{V} = \frac{N}{N_A} \frac{N_A}{V} = n\frac{N_A}{V} = \frac{N_A}{v},
\end{equation}
which leads us to the VDWES,
\begin{equation}
P = \frac{RT}{v-b} -\frac{a}{v^2},
\end{equation}
where the coefficients have been simplified
and redefined as $a=N_A^2a^\prime$ and $b=N_Ab^\prime$.
In addition, the universal gas constant is $R=N_A k$ and the 
molar volume being $v = V/n$.
Although the Lennard–Jones intermolecular potential was used, other potentials that
separate a repulsive region from an attractive one may also be considered. As long as the
corresponding expression can be integrated, the resulting partition function will again lead
to the VDWES.

With the partition function and the second virial coefficient already evaluated, its logarithm
can be written as
\begin{equation}
\ln Q_N\left(T,V\right) = N\ln\left[\left(\frac{V}{N}\right)\left(\frac{2\pi mkT}
{h^2}\right)^{3/2}\right] 
+\ln{\left[1-\frac{N^2}{V}\left(b^\prime -\frac{a^\prime}{kT}\right)\right]}.
\end{equation}
Having established the form of the partition function,
we can now derive the corresponding thermodynamic properties.
The internal energy follows from
\begin{equation}
U =-\left(\frac{\partial}{\partial\beta} \ln Q_N\right)_{V,N} 
= kT^2 \left(\frac{\partial }{\partial T} Q_N\right)_{V,N} 
= \frac{3}{2} NkT -\frac{N}{V} a^\prime .
\end{equation}
To express this result per mole, it is common to introduce the molar internal energy 
$u=U/n$, where $n$ denotes the number of moles. This gives
\begin{equation}
u = \frac{3}{2} RT -\frac{a}{v},
\end{equation}
From the expression for the molar internal energy,
we may directly compute the heat capacity at constant
volume. It is given by
\begin{equation}
c_v = \frac{3}{2} R,
\end{equation}
which is the usual value for the molar heat capacity at
constant volume of a monatomic fluid.
Unlike the ideal gas, however, the internal energy of a 
van der Waals fluid depends not only on temperature
but also on the molar volume.

At fixed temperature, the internal energy increases
as the volume increases. When the
particles form dumbbell-like associations,
they acquire a larger mean free path than free particles,
which contributes to this increase.
Moreover, the internal energy depends linearly
on temperature, so the value of the heat capacity at
constant volume follows directly from the
equipartition theorem, which remains valid for the
translational degrees of freedom of each
monatomic particle.
Building on these thermodynamic relations, we can now determine
the molar entropy and the chemical potential
\begin{eqnarray}
s=\frac{S}{n} &=& \frac{5}{2} R 
+R\ln\left\{\left[\left(\frac{v}{N_A}\right)\left(\frac{2\pi mkT}{h^2}\right)^{3/2}\right]
-\frac{b}{v}\right\}, \label{entropiaVDW1} \\
\mu &=& -kT\left\{\ln\left[\left(\frac{V}{N}\right)\left(\frac{2\pi mkT}{h^2}\right)^{3/2}\right]
-2\frac{N}{V}\left(b^\prime-\frac{a^\prime}{kT}\right)\right\}.
\end{eqnarray}
Note that, except for the last term $-b/v$, the entropy has the same form as that of a
monatomic ideal gas. In this case, the entropy decreases because the structure of the fluid
molecules is known; they are no longer treated as point particles. On the other hand, the
molar Gibbs free energy is given by
\begin{equation}
g = \frac{G}{n} =N_A\mu = 
-RT\ln\left[\left(\frac{v}{N_A}\right)\left(\frac{2\pi mkT}{h^2}\right)^{3/2}\right] 
+2\left(Pv-RT\right).
\end{equation}
In general, it is necessary to describe the VDWF in terms of temperature and pressure,
which are thermodynamic variables that can be controlled experimentally. For this purpose,
the molar Gibbs free energy 
$g$ must be determined.

\subsection{Diatomic fluid}

While it is interesting to understand the thermodynamic behavior of real fluids composed of
monatomic particles, many gases and liquids have more complex molecular structures. In this
section, we analyze a diatomic VDWF. Each molecule is assumed to consist of two atoms
joined by a covalent bond, strong enough that the molecule may be treated as a rigid rotor.
This introduces additional degrees of freedom: three associated with translational motion and
two with rotational motion. Consequently, the corresponding thermodynamic expressions
must be generalized.

We analyze each molecule of an ideal diatomic gas as a rigid rotor. In this approximation,
the Hamiltonian consists solely of the kinetic energy, which can be written as
\begin{equation}
\epsilon = \underbrace{\frac{1}{2M} p_{cm}^2}_{\text{Translational motion}} 
+ \underbrace{\frac{1}{2I} \left[p_{\varphi}^2 
+\frac{1}{\sin^2\left(\varphi\right)}p_{\theta}^2 
\right]}_{\text{Rotational motion}}.
\end{equation}
The linear momentum of the center of mass, denoted by $\vec p_{cm}$,
accounts for the translational motion of the molecule.
This contributes three degrees of freedom, with $M=m_1+m_2$
representing the total mass of the molecule.
In addition, if the bond between the atoms is assumed to be covalent
and therefore very strong, the rotational motion of the molecule
must also be considered.
Under these conditions, the molecule can be modeled as two masses
connected by a rigid bar of length $q$; that is,
as a rigid rotor.
The angular momentum in the azimuthal direction is represented by
$p_{\phi}$ while $p_{\theta}$
corresponds to the angular momentum in the polar direction.
The moment of inertia of the molecule, $I=mq^2$,
is computed using the reduced mass
$m=m_1 m_2 / (m_1+m_2)$.
These rotational degrees of freedom contribute
two additional degrees of freedom for each molecule.
Thus, each molecule has a total of five degrees of freedom:
three associated with translational motion
through the thermodynamic volume, and two associated with rotation.

Given the considerations above, the Hamiltonian of the VDWF is
\cite{pathria, GreinerThermoStatMech}
\begin{equation}
H\left(\vec{q}, \vec{p}\right) = \sum_{i=1}^{3N} \frac{1}{2M} p_{cm\,i}^2 + \sum_{i=1}^N 
\frac{1}{2I} \left[p_{\varphi\,i}^2 
+\frac{1}{\sin^2\left(\varphi\right)}p_{\theta\,i}^2\right]
+\sum_{i<j}\phi_{ij}\left(\vec r_{ij}\right).
\end{equation}
The last sum represents the influence of the intermolecular potential,
which is given by $\phi\left(\vec r_{ij}\right)$,
where $\vec r_{ij} =\vec q_{cm\,j} -\vec q_{cm\,i}$
is the relative coordinate between molecules $i$ and $j$.
Note that, in this case, we only need to consider
the center-of-mass coordinates of each particle;
the corresponding internal relative
coordinates are not included.
With these considerations, we can now apply the multiplicative
property of the partition function which can be written as
\begin{equation} 
Q_N \left(V,T\right) = Q_{R\,N}\left(T\right)Q_{T\,N}\left(V,T\right),
\end{equation}
where
\begin{multline}
Q_{R\,N}\left(T\right)
= \left[\frac{1}{h^2} \int_{-\infty}^\infty \mathrm{e}^{p_\varphi^2/2IkT} 
dp_\varphi \int_0^{2\pi} d\theta 
\int_0^\pi \int_{-\infty}^\infty \mathrm{e}^{-p_\theta^2/2I\sin^2\varphi kT} 
dp_\theta  d\varphi\right]^N \\
=\left[\frac{2Ik}{\hbar^2}T\right]^N
\end{multline}
is the partition function of the rotational motion of the $N$ molecules and
\begin{multline}
Q_{T\,N}\left(V,T\right) = \frac{1}{N!h^{3N}} \int \exp\left[-\beta\left(\sum_{i=1}^{3N} 
\frac{p_{cm\,i}^2}{2M} +\sum_{i<j} \phi_{ij}\left(r_{ij}\right) \right)\right] 
d^{3N}p_{cm}d^{3N}q_{cm} \\
=\frac{1}{N!}\left(\frac{2\pi MkT}{h^2}\right)^{3N/2} 
\int \exp\left[-\beta\sum_{i<j}\phi_{ij}\right] d^{3N}q_{cm},
\end{multline}
is the partition function of the translational part.
This expression contains the configurational integral $Z_N\left(V,T\right)$,
defined in Eq. (\ref{Z_N}), which incorporates the intermolecular potential energy.
To obtain all thermodynamic functions corresponding to the VDWF,
we must expand
$Z_N\left(V,T\right)$
up to the second virial coefficient 
$B\left(T\right)$. Doing so leads to the
following expression for the partition function
\begin{equation}
Q_N\left(V,T\right) = \frac{V^N}{N!}\left[\left(\frac{2\pi MkT}{h^2}\right)^{3/2} 
\left(\frac{2Ik}{\hbar^2}T\right)\right]^N \left[1 -\frac{N^2}{V}\left(b^\prime 
-\frac{a^\prime}{kT}\right)\right].
\end{equation}
This partition function is similar to the one obtained for monatomic particles.
The only difference is the additional factor $Q_{R\,N}\left(T\right)$,
which accounts for the rotational motion.

The thermodynamic functions can be derived from the above
partition function. In particular, using the Helmholtz free
energy allows us to obtain the pressure,
molar entropy, and chemical potential
\begin{eqnarray}
P &=& \frac{RT}{v-b} -\frac{a}{v^2}, \\
s &=& \frac{7}{2} R 
+R\left\{\ln\left[\left(\frac{v}{N_A}\right)\left(\frac{2\pi MkT}{h^2}\right)^{3/2}
\left(\frac{2Ik}{\hbar^2}T\right)\right]-\frac{b}{v}\right\}, \label{entropiaVDW2} \\
\mu &=& kT\left\{ \ln\left[\left(\frac{V}{N}\right)\left(\frac{2\pi MkT}
{h^2}\right)^{3/2}
\left(\frac{2Ik}{\hbar^2}T\right)\right]
+2\frac{N}{V}\left[b^\prime-\frac{a^\prime}{kT}\right]\right\}.
\end{eqnarray}
Consequently, the molar Gibbs free energy in this case is expressed as
\begin{equation}
g = \frac{G}{n} =N_A\mu = 
-RT\ln\left[\left(\frac{v}{N_A}\right)\left(\frac{2\pi mkT}{h^2}\right)^{3/2}
\left(\frac{2Ik}{\hbar^2}T\right)\right] +2\left(Pv-RT\right).
\end{equation}
It is important to note that the VDWES retains the same
functional form in this case, even though the fluid is
composed of diatomic molecules. However, additional terms
must be incorporated into the entropy and chemical
potential to account for rotational motion.
The entropy increases because rotational degrees of freedom
contribute to the overall disorder of the VDWF.
In this case, the internal energy is given by
\begin{equation}
u = \frac{5}{2} RT -\frac{a}{v}
\end{equation}
and, consequently, the heat capacity is
\begin{equation}
c_V =  \frac{5}{2} R.
\end{equation}
This value is expected since each molecule now has
five degrees of freedom.

\section{Liquid-vapor coexistence}

The thermodynamic behavior of the liquid–vapor coexistence
requires that the temperature and pressure of both phases
be equal, thereby defining the saturation temperature 
$T_s$ and saturation pressure $P_s$, respectively.
The Gibbs free energies of the vapor and liquid must also be equal, 
$g_g=g_l$. This condition leads to
\begin{equation}
2P_s\left(v_g-v_l\right) = RT_s\ln\left(\frac{v_g}{v_l}\right), \label{gl.eq.gg}
\end{equation}
which is independent of whether the fluid is
monoatomic or biatomic. Finally, the total 
change in the Gibbs free energies \cite{pathria, GreinerThermoStatMech} of the vapor
and liquid are equal, that is, 
\begin{eqnarray}
dg_g &=& dg_l, \nonumber \\
-s_g dT_s +v_g dP_s &=& -s_l dT_s +v_l dP_s, \nonumber \\
\frac{dP_s}{dT_s} &=& \frac{s_g-s_l}{v_g-v_l} . \label{Clausius-Clapeyron}
\end{eqnarray}
The last expression is known as the Clausius–Clapeyron equation
\cite{fermi, atkins, callen}.
While an equation of state, such as the van der Waals equation,
can describe the states of aggregation in the 
$P$ versus $v$ plane,
the solution of the Clausius–Clapeyron equation defines
the phase boundaries in the 
$P$ versus $T$ plane,
thereby complementing the thermodynamic
description of these phases.
In this sense, solving the Clausius–Clapeyron
equation does not yield an equation of state,
but instead provides additional information
about the transitions between phases.

\begin{figure}[h]
\centering
\includegraphics[width=0.8\textwidth]{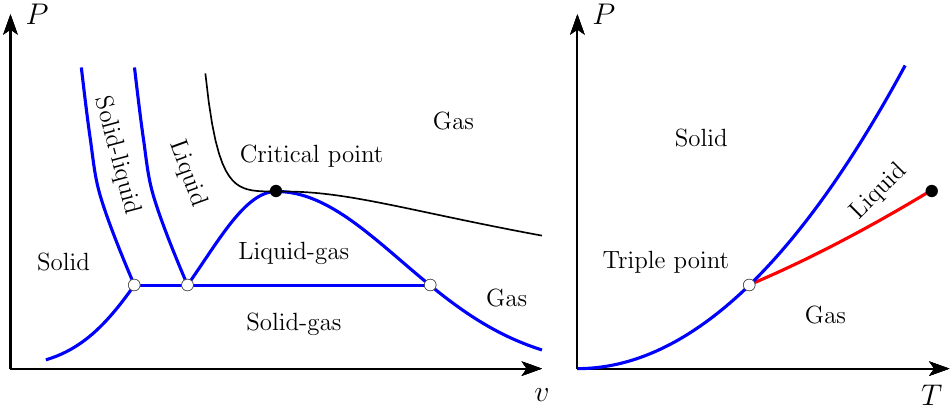}
\caption{
Phase diagrams in the 
$P-v$ and $P-T$ planes,
showing the three states of
aggregation in which a substance composed of fermions,
such as the isotope 
He$^3$ or NO, can exist.
The red curve, which begins at the triple point and ends
at the critical point, represents the liquid–vapor coexistence.
The diagrams are not drawn to scale.}
\label{FD-Phase}
\end{figure}

\begin{figure}[h]
\centering
\includegraphics[width=0.8\textwidth]{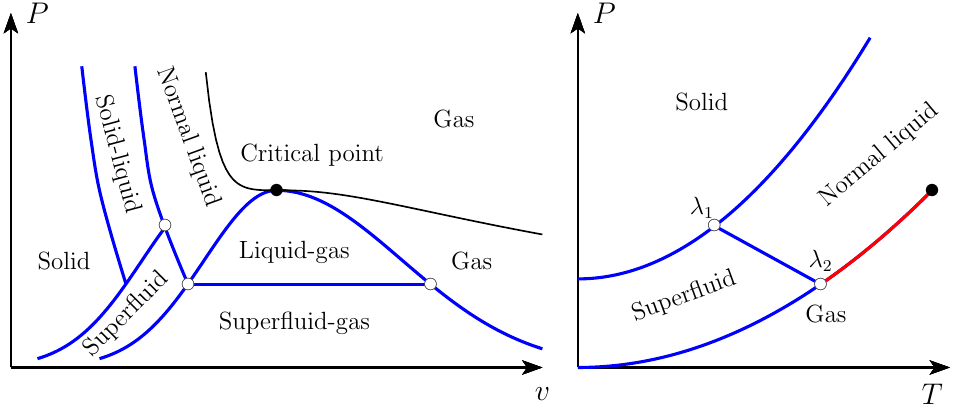}
\caption{
Phase diagrams in the 
$P-v$ and $P-T$ planes,
showing the four states of aggregation in which a substance
composed of bosons, such as the isotope 
He$^4$ or N$_2$, can exist. The superfluid phase is a
macroscopic manifestation of Bose–Einstein condensation.
The diagrams are not drawn to scale.}
\label{BE-Phase}
\end{figure}

For substances whose molecules behave like fermions, according to the Oppenheimer-Ehrenfest 
rule \cite{Oppenheimer1931}, their phase diagrams, $P-v$ and $P-T$, describe three states of aggregation: solid, 
liquid, and gaseous, as shown in Figure \ref{FD-Phase}. In particular, the $P-T$ plane shows 
three coexistence curves that separate two phases and converge at a single point, called 
the triple point because the three phases coexist there. Each curve is the solution to the 
Clausius-Clapeyron equation that describes the coexistence between two phases. The slope 
of this curve depends on the latent heat of transition and the molar volumes of the 
coexisting phases. The fact that $s_B-s_A \neq0$ and $v_B-v_A \neq0$ implies that Gibbs 
free energy is discontinuous when $T=T_s$ and $P=P_s$. This is why these phenomena are 
called first-order transitions.

On the other hand, for substances composed of molecules that behave like bosons, their phase 
diagrams in the $P-v$ and $P-T$ planes now describe four states of matter. That is, in 
addition to solid, liquid, and gas, a superfluid liquid is added (see Figure \ref{BE-Phase}). 
This new phase exhibits, macroscopically, a liquid without viscosity, that is, without 
resistance to movement. This is a manifestation of Bose-Einstein condensation. In the $P-T$ 
plane, five coexistence curves are now plotted since there are five possible transitions: 
solid-superfluid, solid-normal liquid, superfluid-normal liquid, superfluid-gas, and 
liquid-gas. Here, instead of having a triple point, as in the case of fermionic molecules, 
two points have now been formed, $\lambda_1$ and $\lambda_2$, which are the 
limits of the superfluid-normal liquid curve. $\lambda_1$ is a point shared by 
the solid, normal liquid, and superfluid states, while $\lambda_2$ is shared by 
the normal liquid, superfluid, and gaseous states.  Here, the liquid-vapor coexistence curve 
is the one that starts at $\lambda_2$ and goes to the critical point. 

The liquid–vapor coexistence for both types of substances can be described,
with little error, using the thermodynamics and statistics
of a van der Waals fluid by adopting a classical approximation
in the construction of the partition function.
This is justified because the behavior is well captured
near the critical point, though not near the triple point or the 
$\lambda_2$ point.

The most common way to obtain the liquid-vapor coexistence curve is by applying 
Maxwell's rule; see Figure \ref{IsoVanDerWaals}. This rule is used to 
determine the saturation pressure, $p_s$, for an isotherm $t_s$, that is, a 
pressure for which liquid-vapor coexistence is established.
This requires that the areas
\begin{eqnarray}
A_1 &=& \int_{v_{r1}}^{v_{r2}} \left[p_s -\frac{8t_s}{3v_r-1} +\frac{3}{v_r^2}\right]dv_r \\
&=& p_s\left(v_{r2}-v_{r1}\right) -\frac{8}{3}t_s \ln\frac{3v_{r2}-1}{3v_{r1}-1} 
-3\left(\frac{1}{v_{r2}}-\frac{1}{v_{r1}}\right) , \\
A_2 &=& \int_{v_{r2}}^{v_{r3}} \left[\frac{8t_s}{3v_r-1} -\frac{3}{v_r^2} -p_s\right]dv_r \\
&=& \frac{8}{3} t_s \ln\frac{3v_{r3}-1}{3v_{r2}-1} +3\left(\frac{1}{v_{r3}} -\frac{1}{v_{r2}}\right)
-p_s\left(v_{r3}-v_{r2}\right) 
\end{eqnarray}
be equal. Once the saturation pressure is found, with $A_1=A_2$, it is expressed as 
\begin{equation}
p_s = \frac{8}{3} \frac{t_s}{v_{rg}-v_{rl}} \ln\frac{3v_{rg}-1}{3v_{rl}-1} -\frac{3}{v_{rg}v_{rl}}, 
\end{equation}
where $p_s = P_s/P_c$, $t_s=T_s/T_c$, $v_{r\,l} = v_l/v_c$ and $v_{r\,g} = v_g/v_c$. This 
expression is implicit, since the relative volumes depend on both $p_s$ and 
$t_s$. 

The difference in molar entropies that enters the Clausius–Clapeyron
equation can be calculated using either
(\ref{entropiaVDW1}) or (\ref{entropiaVDW2}), and is given by
\begin{equation}
s_g-s_l = R\ln\left(\frac{v_g}{v_l}\right) -\frac{P_s}{T_s} \left(v_g-v_l\right) 
-\frac{a}{T_s} \left(\frac{1}{v_g}-\frac{1}{v_l}\right)
\end{equation}
To simplify this expression, the equality (\ref{gl.eq.gg}) is used,
yielding the Clausius–Clapeyron equation in the form
\begin{equation}
\frac{dP_s}{dT_s} = \frac{P_s}{T_s} +\frac{a}{v_gv_lT_s}. 
\end{equation}
This yields the general form of a coupled first-order
differential equation. The first step is to decouple it.
To do so, it is convenient to express both the van der
Waals and Clausius–Clapeyron equations in terms
of the reduced variables, defined as
\begin{eqnarray}
p_s v_r^3 -\frac{1}{3}\left(8t_s+p_s\right)v_r^2 +3v_r -1=0, \\
\frac{dp_s}{dt_s} = \frac{p_s}{t_s} + 3\frac{1}{v_{rl}v_{rg}t_s}.
\end{eqnarray}
To calculate the relative volumes of the liquid and vapor, it is convenient to define the molar 
volumetric density as $\rho_r = 1/v_r$ and, consequently, rewrite the VDWES as 
\begin{equation}
\rho_r^3-3\rho_r^2+\frac{1}{3}\left(8t_s+p_s\right)\rho_r -p_s =0.
\label{DenVDW}
\end{equation}
This form is particularly advantageous because it makes it easier
to apply Cardano’s method to find its three roots, which satisfy
the relations
$\rho_{r1}\rho_{r2}\rho_{r3} = p_s$ and $\rho_{r1}<\rho_{r2}<\rho_{r3}$.
Here we take 
$v_{rl} = 1/\rho_{r3}$,  $v_{rg} = 1/\rho_{r1}$ and $v_{r2} = 1/\rho_{r2}$.
The Clausius–Clapeyron equation can then be written as
\begin{eqnarray}
\frac{dp_s}{dt_s} &=& \frac{p_s}{t_s} + 3\frac{1}{t_s} \rho_{rg} \rho_{rl} , \nonumber \\
\frac{dp_s}{dt_s} &=& \left(1+3v_{r2}\right)\frac{p_s}{t_s}, \label{ECC:v2}
\end{eqnarray}
That is, everything was written in terms of the second root $v_{r2}$. Figure 
\ref{IsoVanDerWaals} shows that for the isotherm $t_s=0.9$, the saturation pressure 
is $p_s=0.647$ with $v_{r2}\approx1.0$.
As a first approximation, one may therefore take $v_{r2}=1.0$ in general.
With the above considerations, the Clausius–Clapeyron
equation becomes a homogeneous linear differential equation,
whose solution is given by
\begin{equation}
p_s\left(t_s\right) = t_s^4 \label{p_s:0},
\end{equation}
which we will refer to as the zero approximation.
From this point onward, we take the initial condition
$p_s\left(t_s=1\right)=1$ in order to center the solution
at the critical point.
An interesting feature of this approximation is that it reveals
that the slope of the liquid–vapor coexistence curve is large,
suggesting that it can be effectively decoupled from
the Clausius–Clapeyron equation.

\begin{table}
\caption{Comparison between the second
root obtained using Maxwell’s rule and the
one calculated from the expression (\ref{root:v2}).}
\label{Maxwell}
\begin{tabular}{|l|l|l|l|l|}
\hline
\multicolumn{3}{ |c| }{Maxwell's rule} &  
\multicolumn{2}{ |c| }{Estimated}\\
\hline
$t_s$ & $p_s$ & $v_{r\,2}$ & $v_{r\,2}$ & \%err \\
\hline
0.85 & 0.504 & 1.144 & 1.110 & 2.97 \\
0.90 & 0.647 & 1.091 & 1.071 & 1.83 \\
0.95 & 0.812 & 1.043 & 1.034 & 0.86 \\
\hline
\end{tabular}
\end{table}

\begin{figure}
\centering
\includegraphics[width = 0.6\textwidth]{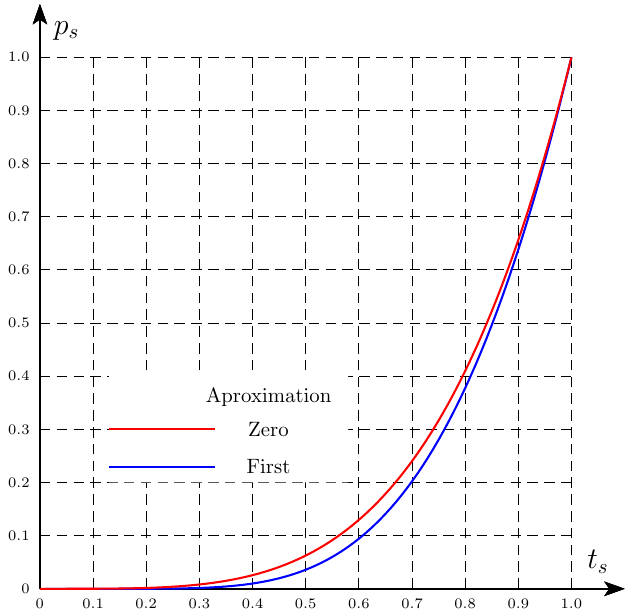}
\caption{Liquid-vapor coexistence curves for the zero and first approximations, where the 
saturation pressure is plotted as a function of temperature in both cases. }
\label{CurvaL-V:0-1} 
\end{figure}

By applying Maxwell’s rule, we observe that $v_{r2}>1$
and that $v_{r2}\to1$ as $t_s\to1$; that is, the second root depends on both 
$t_s$ and $p_s$.
Polynomial (\ref{DenVDW}) can be simplified by eliminating the
quadratic term through the change of variable $\rho_r = w+1$.
This allows us to write the second root, approximately, as follows
\begin{equation}
v_{r\,2} \approx \left[\frac{1}{9}\left(8t_s+p_s\right)\right]^{-1/2} 
= \frac{3}{\left(8t_s+p_s\right)^{1/2}} . \label{root:v2}
\end{equation}
Table \ref{Maxwell} shows three isotherms for which
$p_s$ and $v_{r2}$ are calculated using Maxwell's 
rule. This second root is then compared with the value obtained from the
approximate formula, yielding errors of of $3.00\%$ or less.
With this result, the Clausius–Clapeyron equation becomes
\begin{equation}
\frac{dp_s}{dt_s} = \frac{p_s}{t_s} +\frac{9}{\left(8t_s+p_s\right)^{1/2}} \frac{p_s}{t_s},
\end{equation}
that is, we now have a decoupled differential equation.
Its solution, using the change of variable $u=p_s/t_s$, is given by 
\begin{equation}
\sqrt{8+\frac{p_s}{t_s}} 
+\sqrt{2} \ln \frac{\sqrt{8}-\sqrt{8+p_s/t_s}}{\sqrt{8}+\sqrt{8+p_s/t_s}} = -9t_s^{-1/2} +C_1.
\end{equation}
However, this solution has two drawbacks: it is an implicit function and cannot be solved for 
$p_s$ in terms of $t_s$, and it is not valid at the critical point, when $p_s\left(t_s=1\right)=1$. 
To obtain a viable solution, the value of the second root
can be approximated using a Taylor series, taking into account that 
$8t_s>p_s$
\begin{equation}
v_{r\,2} \approx \frac{3}{2\sqrt{2}t_s^{1/2}} \left(1-\frac{p_s}{16t_s}\right).
\end{equation}
If we take the first term of the Taylor series, we obtain a homogeneous linear differential 
equation 
\[
\frac{dp_s}{dt_s} = \left(\frac{1}{t_s} + \frac{9}{2\sqrt{2}}\frac{1}{t_s^{3/2}}\right)p_s,
\]
whose solution is given by
\begin{equation}
p_s = t_s \exp\left[\frac{9}{\sqrt{2}}\left(1-t_s^{-1/2}\right)\right].
\label{p_s:1}
\end{equation}
We will refer to this result as the first approximation.
Figure \ref{CurvaL-V:0-1} presents the saturation pressure
as a function of temperature for both the zero and first approximations.
As shown, the first approximation exhibits a steeper slope than the zero approximation. 

Another approach consists of retaining both terms of the Taylor expansion.
In this case, the differential equation to be solved becomes
\begin{equation}
\frac{dp_s}{dt_s} = \left(\frac{1}{t_s} + \frac{9}{2\sqrt{2}}\frac{1}{t_s^{3/2}}\right)p_s
-\frac{9}{2^5\sqrt{2}} \frac{p_s^2}{t_s^{5/2}},
\end{equation}
which is now a Bernoulli equation whose solution is
\begin{equation}
p_s = 16t_s \left[15\exp\left[\frac{9}{\sqrt{2}}\left(t_s^{-1/2} -1\right) \right]+1 \right]^{-1}.
\label{p_s:2}
\end{equation}
This is the second approximation. Table \ref{Maxwell:a2} shows three isotherms
for which the saturation pressure is calculated using Maxwell’s rule and the
three approximations derived above. This second approximation exhibits the
smallest error-less than $1\%$—compared to the other two.
Consequently, Maxwell’s rule can be replaced without significantly
affecting the result.
One can then compute the liquid and vapor volumes and fully
describe the liquid–vapor coexistence transition for a VDWF.

\begin{table}
\caption{Comparison among the saturation pressure calculated using Maxwell's rule and those 
obtained by the three approximations.}
\label{Maxwell:a2}
\begin{tabular}{|l|l|l|l|l|l|l|l|}
\hline
\multicolumn{2}{ |c| }{Maxwell's rule} & 
\multicolumn{6}{ |c| }{Estimated values} \\
\hline
$t_s$ & $p_s$ & $p_{s\,0}$ & $\%err$ & $p_{s\,1}$ & $\%err$ & $p_{s\,2}$ & $\%err$ \\
\hline
0.85 & 0.504 & 0.522 & 3.57 & 0.496 & 1.59 & 0.509 & 0.99 \\
0.90 & 0.647 & 0.656 & 1.39 & 0.638 & 1.39 & 0.650 & 0.46 \\
0.95 & 0.811 & 0.815 & 0.49 & 0.805 & 0.74 & 0.813 & 0.25 \\
\hline
\end{tabular}
\end{table}

\section{Conclusions}

In this work we constructed a microscopic description of the van
der Waals equation of state by explicitly evaluating the second virial
coefficient for fluids composed of both monoatomic and diatomic molecules.
Starting from the Lennard–Jones intermolecular potential and employing
the classical canonical ensemble, we derived the configurational
integral up to second order in the virial expansion,
obtaining the van der Waals equation and all its
associated thermodynamic
properties directly from the partition function.
For diatomic molecules, the analysis was generalized by
modeling each particle as a rigid rotor,
which allowed us to incorporate rotational
degrees of freedom through the multiplicative structure of the partition function.
The resulting expressions for the thermodynamic potentials and entropy
fully reproduce those expected from classical thermodynamics,
confirming the consistency of the derivation.

This microscopic framework provides a clear and pedagogical
route to the thermodynamics of van der Waals fluids and enables
a fully analytical treatment of liquid–vapor coexistence.
By solving the Clausius–Clapeyron equation near the critical
point, and by developing increasingly accurate approximations
for the saturation pressure based on the behavior of the
intermediate root of the van der Waals isotherm, we obtained explicit,
closed-form expressions for the coexistence curve. 
The second approximation, in particular,
yields saturation pressures with errors below
one percent when compared with those obtained
through Maxwell’s construction, demonstrating
that the coexistence curve can be reliably computed
without resorting to graphical or numerical methods.

Overall, this study establishes a coherent statistical–mechanical
derivation of the van der Waals thermodynamics and offers
an analytical approach to the description of phase
coexistence that complements and, in some aspects,
surpasses traditional treatments.
The combination of microscopic insight and analytical
accessibility makes these results valuable for both
pedagogical purposes and the deeper understanding of
the liquid–vapor transition in simple fluids.

\section*{Acknowledgements}

This work was supported by DCB UAM-A grant numbers CB003-25.

\bibliography{BiblioArt}

\end{document}